\documentclass[showpacs,preprintnumbers,amsmath,amssymb,floatfix, twocolumn]{revtex4}
\usepackage{graphicx}

\begin{document}

\title{High-order corrections to the holographic entropy bound from quantum field theory}
\date{\today}
   \author{Yi Ling}
    \email{yling@ncu.edu.cn}
    \affiliation{Center for Gravity and Relativistic Astrophysics, Department of Physics, Nanchang University, Nanchang, 330047, China\\
    CCAST (World Laboratory), P.O. Box 8730, Beijing, 100080, China}
    \author{Hongbao Zhang}
    \email{hbzhang@pkuaa.edu.cn}
    \affiliation{Department of Astronomy, Beijing Normal University, Beijing,
100875, China\\
Department of Physics, Beijing Normal University, Beijing, 100875,
China\\
CCAST (World Laboratory), P.O. Box 8730, Beijing, 100080, China}

\begin{abstract}
From the viewpoint of local quantum field theory, this letter
investigates the high-order corrections to the holographic entropy
bound. As a result, the logarithmic correction term appears
naturally with the definite coefficient $-\frac{1}{2}$, which thus
provides another semi-classical platform to examine such
candidates for quantum gravity as Loop Quantum Gravity and String
Theory.
\end{abstract}


\maketitle

In a recent paper\cite{Y}, Yurtsever provides an explanation for
the holographic principle from local quantum field theory. Based
on the cutoff of the one-mode's energy by the Planck scale physics
and the constraint of the total energy of Fock states against
gravitational collapse, Yurtsever shows that the leading term in
the entropy-area relation is linear, which is in full agreement
with the well established holographic entropy bound. However, in
recent years evidence has been mounting that smaller correction
terms may also exist. The purpose of this letter is to further
investigate the high-order corrections to the holographic entropy
bound from the viewpoint of local quantum field theory.

To proceed, we shall confine ourselves to the real massless scalar
in the spherical region with the radius $R$, which is probably the
more natural choice than the cube considered by Yurtsever. In
addition, as is done by Yurtsever, we assume that the maximum
energy of admissible one-modes is of the order of the Planck
energy, i.e. given by $\Lambda$, where $\Lambda$ is of the order
of one for naive Planck scale physics. Moreover, the maximum
energy of a state in the symmetric Fock space should be
constrained by the energy of a Schwarzschild black hole with
radius R, i.e. $\frac{R}{2}$.

Then by the simplified derivation of Yurtsever's result\cite{A}, the
dimension of the gravitational truncated symmetric Fock space
approximately reads
\begin{equation}
W=I_0(2\sqrt{z}).
\end{equation}
Here $I_0$ is the zeroth-order Bessel function of the second kind,
and $z$ is obtained by
\begin{equation}
z=\int_0^\Lambda
d\omega\rho(\omega)\frac{R}{2\omega}=\frac{\Lambda^2}{6\pi\cdot(4\pi)^2}A^2,
\end{equation}
where the density of states
$\rho(\omega)=\frac{V}{2\pi^2}\omega^2$ has been employed with the
volume $V=\frac{4\pi}{3}R^3$ and the area $A=4\pi R^2$. Later,
according to the asymptotic behavior of $I_0$ as
$x\rightarrow\infty$, i.e.,
\begin{equation}
I_0(x)\sim\frac{e^x}{\sqrt{2\pi x}},
\end{equation}
the maximum entropy takes the form
\begin{equation}
S_{max}=\ln W\sim
\frac{2\Lambda}{4\pi\cdot\sqrt{6\pi}}A-\frac{1}{2}\ln A.
\end{equation}
Some remarks on this result are presented in order. Firstly the
leading term is obviously dependent on the the Planck scale
physics where the ordinary quantum field theory is broken.
Specifically, if $\Lambda=\frac{\pi\cdot\sqrt{6\pi}}{2}$, the
leading term is the just the famous linear entropy-area relation
formula. While the sub-leading correction is independent of the
Planck scale physics. In particular, it is worth noting that the
logarithmic term has also arisen in both Loop Quantum Gravity and
String Theory. However, one should emphasize that there is no
general agreement on the coefficient of the logarithmic
correction\cite{H}. Therefore the resultant coefficient
$-\frac{1}{2}$ obtained here acquires much importance: since our
foregoing assumptions only rely on classical general relativity
and quantum field theory below the Planck scale, it provides
another semi-classical constraint onto all candidates for quantum
gravity such as Loop Quantum Gravity and String Theory.

In summary, based on the viewpoint of local quantum field theory,
which was initially proposed by Yurtsever to explain the famous
linear entropy-area relation, we have shown that the logarithmic
correction to the holographic entropy naturally appears with the
definite coefficient $-\frac{1}{2}$, which should serve as another
severe semi-classical constraint onto any candidate for quantum
gravity.
\section*{Acknowledgements}
Y. Ling's work is partly supported by NSFC (Nos.10405027 and
10205002) and SRF for ROCS, SEM. H. Zhang's work is supported in
part by NSFC(Nos.10373003 and 10533010).

\end{document}